\pdfoutput=1

\documentclass[sigconf,screen]{acmart}
\acmPrice{}
\acmDOI{10.1145/3395363.3397385}
\acmYear{2020}
\copyrightyear{2020}
\acmSubmissionID{issta20main-id58-p}
\acmISBN{978-1-4503-8008-9/20/07}
\acmConference[ISSTA '20]{Proceedings of the 29th ACM SIGSOFT International Symposium  on  Software Testing and Analysis}{July 18--22, 2020}{Los Angeles, CA, USA}
\acmBooktitle{Proceedings of the 29th ACM SIGSOFT International Symposium  on  Software Testing and Analysis (ISSTA '20), July 18--22, 2020, Los Angeles, CA, USA}
\pdfoutput=1

\usepackage{pifont}
\usepackage{csquotes}
\usepackage{xspace}
\usepackage{booktabs}
\usepackage{listings, xcolor}
\usepackage{algpseudocode,algorithm}
\usepackage{tabularx,colortbl}
\acmBadgeR[https://www.acm.org/publications/policies/artifact-review-badging]{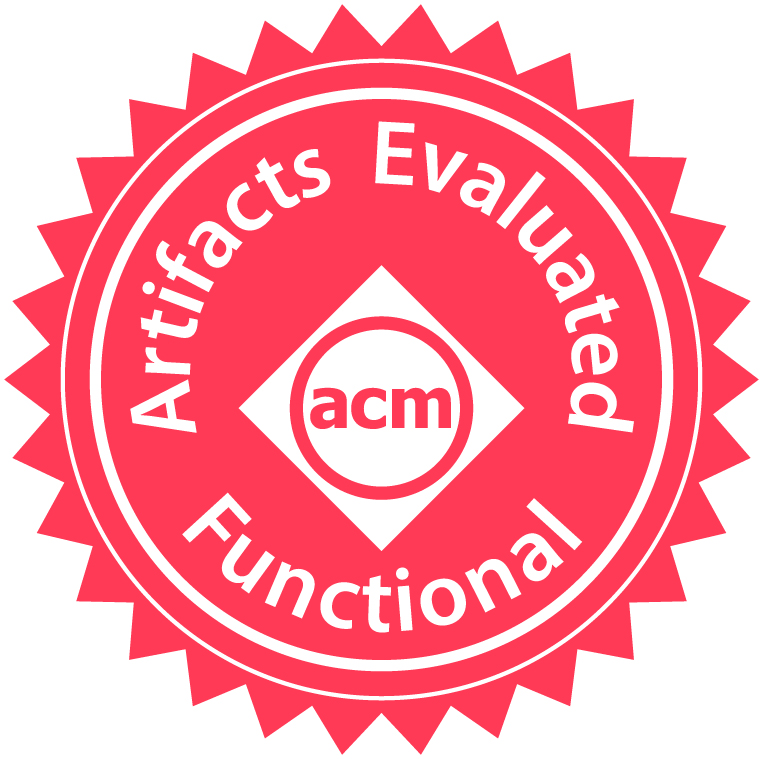}

\usepackage{comment}

\definecolor{verylightgray}{rgb}{.97,.97,.97}

\definecolor{asparagus}{rgb}{0.53, 0.66, 0.42}

\long\def\authornote#1{%
  \leavevmode\unskip\raisebox{-3.5pt}{\rlap{$\scriptstyle\diamond$}}%
  \marginpar{\raggedright\hbadness=10000
    \def\baselinestretch{0.8}\tiny
    \it #1\par}}

\newcommand{\sysname}{{\em SolidiFI}\xspace}

\lstdefinelanguage{Solidity}{
	keywords=[1]{anonymous, assembly, assert, balance, break, call, callcode, case, catch, class, constant, continue, constructor, contract, debugger, default, delegatecall, delete, do, else, emit, event, experimental, export, external, false, finally, for, function, gas, if, implements, import, in, indexed, instanceof, interface, internal, is, length, library, log0, log1, log2, log3, log4, memory, modifier, new, payable, pragma, private, protected, public, pure, push, require, return, returns, revert, selfdestruct, send, solidity, storage, struct, suicide, super, switch, then, this, throw, transfer, true, try, typeof, using, value, view, while, with, addmod, ecrecover, keccak256, mulmod, ripemd160, sha256, sha3}, 
	keywordstyle=[1]\color{blue}\bfseries,
	keywords=[2]{address, bool, byte, bytes, bytes1, bytes2, bytes3, bytes4, bytes5, bytes6, bytes7, bytes8, bytes9, bytes10, bytes11, bytes12, bytes13, bytes14, bytes15, bytes16, bytes17, bytes18, bytes19, bytes20, bytes21, bytes22, bytes23, bytes24, bytes25, bytes26, bytes27, bytes28, bytes29, bytes30, bytes31, bytes32, enum, int, int8, int16, int24, int32, int40, int48, int56, int64, int72, int80, int88, int96, int104, int112, int120, int128, int136, int144, int152, int160, int168, int176, int184, int192, int200, int208, int216, int224, int232, int240, int248, int256, mapping, string, uint, uint8, uint16, uint24, uint32, uint40, uint48, uint56, uint64, uint72, uint80, uint88, uint96, uint104, uint112, uint120, uint128, uint136, uint144, uint152, uint160, uint168, uint176, uint184, uint192, uint200, uint208, uint216, uint224, uint232, uint240, uint248, uint256, var, void, ether, finney, szabo, wei, days, hours, minutes, seconds, weeks, years},	
	keywordstyle=[2]\color{teal}\bfseries,
	keywords=[3]{block, blockhash, coinbase, difficulty, gaslimit, number, timestamp, msg, data, gas, sender, sig, value, now, tx, gasprice, origin},	
	keywordstyle=[3]\color{violet}\bfseries,
	identifierstyle=\color{black},
	sensitive=false,
	comment=[l]{//},
	morecomment=[s]{/*}{*/},
	commentstyle=\color{gray}\ttfamily,
	stringstyle=\color{red}\ttfamily,
	morestring=[b]',
	morestring=[b]",
	basicstyle={\footnotesize \ttfamily}
}

\AtBeginDocument{%
  \providecommand\BibTeX{{%
    \normalfont B\kern-0.5em{\scshape i\kern-0.25em b}\kern-0.8em\TeX}}}

\copyrightyear{2020}
\acmYear{2020}
\setcopyright{acmcopyright}\acmConference[ISSTA '20]{Proceedings of the 29th ACM
SIGSOFT International Symposium on Software Testing and Analysis}{July 18--22,
2020}{Los Angeles/Virtual, CA, USA}
\acmBooktitle{Proceedings of the 29th ACM SIGSOFT International Symposium on
Software Testing and Analysis (ISSTA '20), July 18--22, 2020, Los Angeles/Virtual,
CA, USA}
\acmPrice{15.00}
\acmDOI{10.1145/3395363.3397385}
\acmISBN{978-1-4503-8008-9/20/07}

\begin{document}

\title{How Effective are Smart Contract Analysis Tools?  Evaluating Smart Contract Static Analysis Tools Using Bug Injection}

\author{Asem Ghaleb}
\email{aghaleb@alumni.ubc.ca}
\affiliation{%
  \institution{University of British Columbia}
  \streetaddress{}
  \city{Vancouver}
  \state{}
   \country{Canada}
}
\author{Karthik Pattabiraman}
\email{karthikp@ece.ubc.ca}
\affiliation{%
  \institution{University of British Columbia}
  \streetaddress{}
  \city{Vancouver}
  \state{}
 \country{Canada}
 }

\begin{abstract}
 \pdfoutput=1

Security attacks targeting smart contracts have been on the rise, which have led to financial loss and erosion of trust.
Therefore, it is important to enable developers to discover security vulnerabilities in smart contracts before deployment. 
A number of static analysis tools have been developed for finding security bugs in smart contracts. However, despite the numerous bug-finding tools, there is no systematic approach to evaluate the proposed tools and gauge their effectiveness. 
This paper proposes \sysname, an automated and systematic approach for evaluating smart contracts' static analysis tools. \sysname is based on injecting bugs (i.e., code defects) into all potential locations in a smart contract to introduce targeted security vulnerabilities. \sysname then checks the generated buggy contract using the static analysis tools, and identifies the bugs that the tools are unable to detect (false-negatives) along with identifying the bugs reported as false-positives. \sysname is used to evaluate six widely-used static analysis tools, namely, Oyente, Securify, Mythril, SmartCheck, Manticore and Slither,  using a set of 50 contracts injected by 9369 distinct bugs. It finds several instances of bugs that are not detected by the evaluated tools despite their claims of being able to detect such bugs, and all the tools report many false positives. 

\end{abstract}

\ccsdesc[500]{Security and privacy}
\ccsdesc[300]{Security and privacy~Software and application security}

\keywords{Ethereum, Ethereum security, solidity code analysis, smart contracts, smart contracts security, smart contracts analysis, smart contracts dataset, static analysis tools evaluation, bug injection, fault injection}

\maketitle

\pdfoutput=1

\section{Introduction}

The past few years have witnessed a dramatic rise in the popularity of smart contracts \cite{clack2016smart}. Smart contracts are small programs written into blocks running on top of a blockchain that can receive and execute transactions autonomously without trusted third parties~\cite{grishchenko2018semantic}.
Ethereum  \cite{buterin2014ethereum} is the most popular framework for executing smart contracts. 


Like all software, smart contracts may contain bugs. 
Unfortunately, bugs in smart contracts can be exploited by malicious attackers for financial gains. 
In addition, transactions on Ethereum are immutable and cannot be reverted, so losses cannot be recovered. 
Further, it is difficult to update a smart contract after its deployment. Consequently, there have been many bugs in smart contracts that have been maliciously exploited in the recent past \cite{dao2017, paritywalletbreach2017, mathieu2017blocktix}.
Therefore, there is a compelling need to analyze smart contracts to detect and fix security bugs.

Several approaches and tools have been developed that statically find security bugs in smart contracts~\cite{luu2016making, mueller2018smashing, tsankov2018securify, tikhomirov2018smartcheck, feist2019slither}. 
However, despite the prevalence of these static analysis tools, security bugs abound in smart contracts~\cite{perez2019smart}. This calls into
question the efficacy of these tools and their associated techniques. Unfortunately, many of the static analysis tools
have been evaluated either only by their developers on custom data-sets and inputs, often in an ad-hoc manner, or on data-sets of contracts with a limited number of bugs (112 bugs \cite{durieux2019empirical} and 10 bugs \cite{parizi2018empirical}). {\em To the best of our knowledge,
there is no systematic method to evaluate static analysis tools for smart contracts regarding their effectiveness in finding security bugs.} 

Typically, static analysis tools can have both false-positives and false-negatives. While false positives are important,  false negatives in smart contracts can lead to critical consequences, as exploiting bugs in contracts usually leads to loss of ether (money). Also, empirical studies of software defects in the field have found that many of the defects can be detected by static analysis tools in theory, but are not detected due to limitations of the tools~\cite{thung2012extent}. 
In our work, we focus mostly on the undetected bugs (i.e., false negatives), though we also study false-positives of the tools.

We perform bug injection to evaluate the false-negatives of smart contract static analysis tools. 
Bug injection as a testing approach has been extensively explored in the domain of traditional programs \cite{pewny2016evilcoder,dolan2016lava,bonett2018discovering}; 
however, there have been few papers on bug injection in the context of smart contracts. 
This problem is challenging for two reasons. 
First, smart contracts on Ethereum are written using the Solidity language, which differs from conventional programming languages 
typically targeted by mutation testing tools \cite{dannen2017introducing}.
Second, because our goal is to inject security bugs, the bugs injected should lead to exploitable vulnerabilities. 

This paper proposes \sysname\footnote{\sysname stands for Solidity Fault Injector, pronounced as Solidify. }, a methodology for systematic evaluation of smart contracts' 
static analysis tools to discover potential flaws in the tools that lead to undetected security bugs. 
\sysname injects bugs formulated as code snippets into {\em all} possible locations into a smart contract's source code written in Solidity. 
The code snippets are vulnerable to specific security vulnerabilities that can be exploited by an attacker. 
The resulting buggy smart contracts are then analyzed using the static analysis tools being evaluated, and the results are inspected for those injected bugs that are not detected by each tool - these are the false-negatives of the tool. 
Because our methodology is agnostic of the tool being evaluated, it can be applied to any static analysis tool that works on Solidity.  

We make the following contributions in this paper.

\begin{itemize}	
    \item Design a systematic approach for evaluating false-negatives and false-positives of smart contracts' static analysis tools.
    \item Implement our approach as an automated tool, \sysname, to inject security bugs into smart contracts written in Solidity.    
    \item Use \sysname to evaluate six static analysis tools of Ethereum smart contracts for false-negatives and false-positives.
     \item Provide an analysis of the undetected security bugs and false-positives for the 6 tools, and the reasons behind them. 
     \end{itemize}

The results of using \sysname on 50 contracts show that all of the evaluated tools had significant false-negatives ranging from $129$ to $4137$ undetected bugs across 7 different bug types despite their claims of being able to detect such bugs, as well as many false positives. Further, many of the undetected bugs were found to be exploitable when the contract is executed on the blockchain.
Finally, we find that \sysname takes less than 1 minute to inject bugs into a smart contract (on average). 
Our results can be used by tool developers to enhance the evaluated tools, and by researchers proposing new bug-finding tools for smart contracts. 


\pdfoutput=1

\section{Background}
\label{sec:background}


\subsection{Smart Contracts}
%
As mentioned earlier, smart contracts are written in a high-level language such as Solidity. They are compiled to Ethereum Virtual Machine (EVM) bytecode that is deployed and stored in the blockchain accounts. Smart contract transactions are executed by miners, which are a network of mutually untrusted nodes, and governed by the consensus protocol of the blockchain. Miners receive execution fees, called gas, for running the transactions which are paid by the users who submit the execution requests. We illustrate smart contracts through a running example shown in Figure~\ref{fig:example-contract} (adapted from prior work~\cite{atzei2017survey}). 

\begin{figure}[h]
\footnotesize
\centering
\begin{lstlisting}[language= Solidity, basicstyle=\ttfamily]
pragma solidity >=0.4.21 <0.6.0;
contract EGame{
    address payable private winner;
    uint startTime;
    
    constructor() public{ 
      winner = msg.sender;
      startTime = block.timestamp;}
    
    function play(bytes32 guess) public {
     if(keccak256(abi.encode(guess)) == keccak256(abi.encode('solution'))){
        if (startTime + (5 * 1 days) == block.timestamp){
           winner = msg.sender;}}}
           
    function getReward() payable public{
      winner.transfer(msg.value);}
 }    
\end{lstlisting}
\caption{Simple contract written in Solidity.}
\label{fig:example-contract}
\end{figure}

This contract implements a public game that enables users to play a game and submit their guesses or solutions for the game along with some amount of money. The money will be transferred to the account of the last winner if the guess is wrong; otherwise the user will be set as the current winner and will receive the money from users who play later. The \textit{constructor()} at line 6 runs only once when the contract is created, and it sets the initial winner to the owner of the contract defined by the user who submitted the create transaction of the contract (\textit{msg.sender}). It also initializes the \textit{startTime} variable to the current timestamp during the contract creation. The function \textit{play} at line 10 is called by the user who wants to submit his/her guess, and it compares the received guess with the true guess value. If the comparison is successful, it sets the \textit{winner} to the address of the user account who called this function, provided the guess was submitted within 5 days of creating the contract. Finally, the function \textit{getReward} sends the amount of \textit{ether} specified in the call to getReward (\textit{msg.value)}, to the last winner.

\subsection{Static Analysis Tools}
We consider six static analysis tools for finding bugs in smart contracts in this paper, Oyente, Securify, Mythril, Smartcheck, Manticore, and Slither. 
They all operate on smart contracts written in Solidity, and are freely available. Further, they are all automated and require no annotations from the programmer.  
We selected Oyente \cite{luu2016making}, Securify \cite{tsankov2018securify}, Mythril \cite{mueller2018smashing}, and Manticore \cite{mossberg2019manticore} as they were used in many smart contract analysis studies \cite{tsankov2018securify, perez2019smart, parizi2018empirical, brent2018vandal}. We included Smartcheck \cite{tikhomirov2018smartcheck} as it uses a pattern matching approach rather than symbolic execution employed by the previous four tools. Similarly Slither \cite{feist2019slither} is another non-symbolic-execution based tool, but unlike SmartCheck, it uses Static Single Assignment (SSA) for analysis.  

\pdfoutput=1

\section{Motivation and Challenges}
\label{sec:vulnerabilities}

This section first presents motivating examples of undetected security bugs by static analysis tools, followed by an overview of the challenges in the evaluation of the tools.

\subsection{Motivating examples}

The contract example in Figure \ref{fig:example-contract} has at least 2  vulnerabilities, (1) two instances of timestamp dependency bug at lines 8 and 12, and (2) one instance of transaction ordering dependence (TOD) represented by the transactions at lines 13 and 16. 
The timestamp dependency bug is that the block's timestamp should not be used in the transaction, while the TOD bug is that the state of the smart contract should not be relied upon by the developer (Section~\ref{sec:eva}).

We have used four of the static analysis tools in Section~\ref{sec:background} (supposed to detect these bugs) to check this contract for bugs, Oyente, Securify, Mythril, and SmartCheck. According to the tools' research papers \cite{luu2016making, tsankov2018securify, mueller2018smashing, tikhomirov2018smartcheck}, Oyente, Mythril and SmartCheck should detect the timestamp dependency bug. 
However, we found that while Oyente and Mythrill were not able to detect both instances of timestamp dependency bug in lines 8 and 12,  Smartcheck detected only the instance in line 12. For the second instance, Smartcheck gave a hint that block.timestamp should be ``used only in equalities''. To further test SmartCheck's ability to detect the bug, we made a small modification to the syntax of the smart contract while keeping its semantics the same (Figure~\ref{fig:contract-modification}). SmartCheck subsequently failed to detect the bug altogether.

\begin{figure}
\footnotesize
\centering
\begin{lstlisting}[firstnumber=11]
  uint _vtime = block.timestamp;
  if (startTime + (5 * 1 days) == _vtime){
\end{lstlisting}
\caption{Modification made to the contract in Figure \ref{fig:example-contract}}
\label{fig:contract-modification}
\vspace{-3mm}
\end{figure}

Regarding the TOD bug, both Oyente and Securify are supposed to detect this class of bugs. 
However, we found that only Securify detected this bug successfully while Oyente was not able to detect it.
We extracted the code snippet representing TOD from this contract (lines 10 to 16), injected it in another larger contract free of bugs, and obtained similar results. 

These examples motivated us to prepare multiple code snippets for the different bugs (within the scope of the tools) and to manually inject them into the code of 5 smart contracts (the first 5 contracts in the set of contracts in Section~\ref{sec:eva}). We then used the tools to check the buggy contracts, and found several instances of undetected bugs even though the tools were supposed to detect them. However, it was tedious and error-prone to manually inject these bugs and inspect the results, and so we decided to automate this process. 

\subsection{Automated bug injection challenges}
\label{injectionchallenges}

The simplest way to inject bugs into smart contracts is to inject them at random locations - this is how traditional fault injection (i.e., mutation testing) works. However, random injection is not a cost-effective approach as we have to follow specific guidelines for the injected bug to be exploitable. We identify two main challenges.

\subsubsection{Bug injection locations}
As the underlying techniques used by some tools (e.g., symbolic execution) depends on the control and data flow in the analyzed contracts, injecting an instance of each bug at a single location would not be sufficient. Therefore, bugs should be injected into all potential locations in the contract code. On the other hand, the process of identifying the potential locations depends on the code of the original contract, and also on the type and nature of each bug. Injecting bugs at the wrong locations would result in compilation errors. In addition, it might yield instances of dead code in the contract. For example, injecting a bug formulated as a stand-alone function inside the body of another function would result in a compilation error, as Solidity does not support nested functions. 
Moreover, a bug injected into an 'if' statement condition  that would make the condition always fail, would make the 'then' clause unreachable.
 
\subsubsection{Semantics dependency}
For the injected bug to be an active bug that can be exploited by an attacker, it has to be aligned with the semantics of the original contract. 
For example, assume that we want to inject a Denial of Service (DoS) bug by calling an external contract. We can use an if-statement with a condition containing a call to another contract function. However, for this bug to be executed, we also need to define the appropriate external contract.


\sysname addresses the first challenge by parsing the Solidity language into an Abstract Syntax Tree (AST) and injecting bugs into {\em all} syntactically valid locations. It addresses the second challenge by formulating exploitable code snippets for each bug type.
\pdfoutput=1

\section{\sysname Approach and Workflow}
\label{sec:SolidiFI}

\begin{figure}[h] 
\centering
\includegraphics[width=\columnwidth]{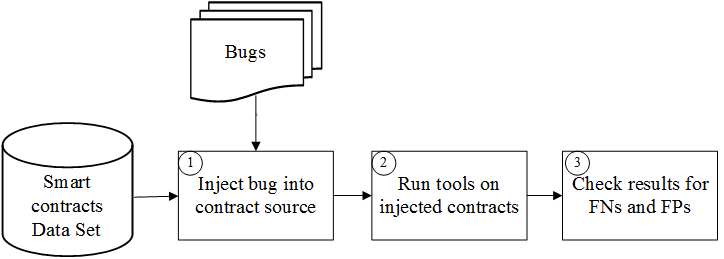}
\caption{\sysname Workflow.}
\label{gprocess}
\end{figure}

The main goal of \sysname is to perform systematic evaluation of \textit{static analysis} tools used to check smart contracts for known security bugs. Figure \ref{gprocess} shows the workflow of \sysname. 
The code snippets representing a specific security bug are injected in each smart contract's source code at {\em all} possible locations (step 1). The selection of the injection locations is a function of the bug to be injected. {\em \sysname injects bugs into the source code to imitate the introduction of bugs by developers. However, its use is not restricted to tools that perform analysis at the source code level. For example, tools that work on the EVM bytecode would compile the buggy contracts to produce the EVM code for analysis.} Then, the injected code is scanned using the static analysis tools (step 2). Finally, the results of each tool are checked, and false negatives and false positives are measured (step 3).


\subsection{Bug Model}

In our work, a security bug is expressed as a code snippet, which leads to a vulnerability that the security tool being analyzed aims to detect. \sysname reads code snippets to be injected from a pre-defined bug pool prepared by us (the bug pool can be easily extended by users to add new bugs). For each tool, we only inject the bugs that the tool claims to detect. based on the tool's research paper. 
However, because the tools are continuously evolving,  the research paper may not have the up-to-date list of bugs detected by the tool, and hence we use the tool's online documentation to augment it. 


\subsection{Bug Injection}
\label{buginstrument}
In this work, the security bugs are injected in the source code in three ways as follows.

\subsubsection{Full code snippet}
In this approach, we prepare several code snippets  for each bug under study. Each code snippet is a piece of code that introduces the security bug. 
To illustrate the process, we discuss the bugs and example code snippets. 

\paragraph{\textbf{Timestamp dependency}}
The current timestamp of the block can be used by contracts to trigger some time-dependent events. Given the decentralized nature of Ethereum, 
miners can change the timestamp (to some extent). Malicious miners can use this capability and change the timestamp to favor themselves. This bug was exploited in the GoverMental Ponzi scheme attack \cite{EthereumAttacksHistory}. Therefore, developers should not rely on the precision of the block's timestamp. Figure~\ref{fig:timestamp-example} shows an example of a 
code snippet that represents  the bug (\textit{block.timestamp} returns the block's timestamp).

\begin{figure}[h]
\centering
\footnotesize
\begin{lstlisting}[language= Solidity, basicstyle=\ttfamily]
function bug_tmstmp() public returns(bool)
{   return block.timestamp >= 1546300;}
\end{lstlisting}
\caption{Timestamp dependency examples.}
\label{fig:timestamp-example}
\end{figure}

\paragraph{ \textbf{Unhandled exceptions}}
In Ethereum, contracts can call each other, and send \textit{ether} to each other (e.g., \textit{send} instruction, \textit{call} instruction, etc.). If an exception is thrown by the callee contact (e.g., limited gas for execution), the contract is terminated, its state is reverted, and  \textit{false} is returned to the caller contract. Therefore, unchecked returned values within the caller contract could be used to attack the contract, leading to undesired behavior. A serious version of this bug occurred in the ``King of the Ether" \cite{EthereumAttacksHistory}. Figure~\ref{fig:unhandled-exception-example} shows an example (the \textit{send()} instruction requires its return value to be checked for exceptions to make it secure).
\begin{figure}[h]
\footnotesize
\centering
\begin{lstlisting}[basicstyle=\ttfamily]
function unhandledsend() public {
    callee.send(5 ether);}
\end{lstlisting}
\caption{Unhandled exceptions examples.}
\label{fig:unhandled-exception-example}
\end{figure}
 \paragraph {\bf Integer overflow/underflow} In Solidity, storing a value in an integer variable bigger or smaller than its limits lead to integer overflow or underflow. This can be used by attackers to fraudulently siphon off funds. For example, Figure \ref{fig:int-overflow-example} shows an example code snippet in which an attacker can reset the {\em lockTime} for a user by calling the function \textit{incrLockTime} and passing 256 as an argument - this would cause an overflow, and end up  setting the lockTime to 0. Batch Transfer Overflow is a real-world example \cite{PoWHC}.
 \begin{figure}[h]
 \footnotesize
\centering
\begin{lstlisting}[basicstyle=\ttfamily]
function incrLockTime(uint _sec) public{
        lockTime[msg.sender] += _sec;}
\end{lstlisting}
\caption{Integer overflow/underflow example.}
\label{fig:int-overflow-example}
\end{figure}

\paragraph {\bf Use of tx.origin} In a chain of calls, when contracts call functions of each other, the use of  {\em tx.origin} (that returns the first caller that originally sent the call) for authentication instead of {\em msg.sender} (that returns the immediate caller) can lead to phishing-like attacks \cite{solidity-security-blog}. Figure \ref{fig:tx.origin} shows an example snippet in which {\em tx.origin} is used to withdraw money. 
\begin{figure}[h]
\footnotesize
\centering
\begin{lstlisting}[basicstyle=\ttfamily]
function bug_txorigin(address _recipient) public {
    require(tx.origin == owner);
    _recipient.transfer(this.balance);}
\end{lstlisting}
\caption{tx.origin authentication example.}
\label{fig:tx.origin}
\end{figure}

 \paragraph{\bf Re-entrancy} Contracts expose external calls in their interface. These external calls can be hijacked by attackers to call a function within the contract itself several times, thereby performing unexpected operations within the contract itself. For example, the external call in Line 3 of the snippet code shown in Figure \ref{fig:Re-entrancy} can be used by an attacker to call the {\em bug\_reEntrancy()} function repeatedly, potentially leading to withdrawal of ether more than the balance of the user. The DAO attack \cite{dao2017} is a well-known example exploiting this bug.
 \begin{figure}[h]
 \footnotesize
\centering
\begin{lstlisting}[basicstyle=\ttfamily]
function bug_reEntrancy(uint256 _Amt) public {
  require(balances[msg.sender] >= _Amt);
  require(msg.sender.call.value(_Amt));
  balances[msg.sender] -= _Amt;}
\end{lstlisting}
\caption{Re-entrancy example.}
\label{fig:Re-entrancy}
\end{figure}
\paragraph{\bf Unchecked send} Unauthorized Ether transfer, such as non zero sends, can be called by external users if they are visible to public, even if they do not have the correct credentials. This means unauthorized users can call such functions and transfer ether from the vulnerable contract \cite{solidity-security-blog}. An example code snippet is shown in Figure~\ref{fig:uncksend}.

\begin{figure}[h]
\footnotesize
\centering
\begin{lstlisting}[basicstyle=\ttfamily]
function bug_unchkSend() payable public{
      msg.sender.transfer(1 ether);}
\end{lstlisting}
\caption{Unchecked send example.}
\label{fig:uncksend}
\end{figure}

\paragraph {\bf Transaction Ordering Dependancy (TOD)} Changing the order of the transactions in a single block that has multiple calls to the contract, results in changing the final output \cite{atzei2017survey}. Malicious miners can benefit from this. An example code snippet vulnerable to this bug is shown in Figure~\ref{fig:tod}. In this example, the attackers can send a puzzle solving reward to themselves instead of the winner of the game by executing {\em bug\_tod2()} before {\em bug\_tod1()}.
\begin{figure}[h]
\footnotesize
\centering
\begin{lstlisting}[language= Solidity, basicstyle=\ttfamily]
address payable winner_tod;
function setWinner_tod() public {
    winner_tod = msg.sender;}
function getReward_tod() payable public{
    winner_tod.transfer(msg.value);}
\end{lstlisting}
\caption{TOD example.}
\label{fig:tod}
\end{figure}



\subsubsection{Code transformation}
This approach aims to transform a piece of code without changing its functionality, but make it vulnerable to a specific bug. We leverage known patterns of vulnerable code to inject this bug. We use this approach to inject two bug classes that are compatible with this approach, namely (1) integer overflow/underflow and (2) use of tx.origin. 

Table \ref{tab:codetranformpattern} shows examples of the code patterns that are replaced to introduce the bugs, and the vulnerable patterns for each bug type.

\begin{table}[h]
\begin{center}
\caption{Code transformation patterns.}
\begin{tabular}{|c|c|c|}  
\hline
 \textbf{Bug Type}&{\textbf{Original Code Patterns}}&{{\textbf{New Code Patterns}}} \\
\hline
tx.origin&msg.sender==owner& tx.origin==owner\\
\hline
Overflow&bytes32&bytes8\\
\hline
Overflow&uint256&uint8\\
\hline

\end{tabular}
\label{tab:codetranformpattern}
\end{center}
\end{table}

Figure \ref{fig:code_trans_example}  shows an example before and after bug injection using this approach. In this example, \textit{transfer} instruction is used to perform a transfer of the specified ether amount to the receiver's account after verifying the direct caller of \textit{sendto()} to be the owner. To inject the tx.origin bug, the authorization condition \textit{msg.sender == owner} should be replaced with the \textit{tx.origin == owner}, in which the owner is not the direct caller of \textit{sendto()}. However, the authorization check is passed successfully, which enables attackers to authorize themselves, and send ether from the contract, even if they are not the owner.

\begin{figure}[h]
\centering
\footnotesize
\begin{lstlisting}[basicstyle=\ttfamily]
/*(Before)*/
function sendto(address receiver, uint amount) public  {
    require (msg.sender == owner);
    receiver.transfer(amount);}
/*(After injection)*/
function sendto(address receiver, uint amount) public {
    require (tx.origin == owner);
    receiver.transfer(amount);}
  \end{lstlisting}
 \caption{Code transformation example.}
 \label{fig:code_trans_example}
\end{figure}

\subsubsection{Weakening security mechanisms} 
In this approach, we weaken the security protection mechanisms in the smart contract code, which protect external calls. Note that our goal is to evaluate the static analysis tool, and not the smart contract itself. 
 We use this approach to inject Unhandled exception bugs. 
 Figure~\ref{fig:weakening-example} shows an example, in which the  \textit{Unhandled exceptions} bug is injected by 
 removing the {\em revert()} statement that reverts the state of the contract if the transfer transaction failed - this causes the balance to incorrectly become 0 even if the transaction failed.

\begin{figure}[h]
\centering
\footnotesize
\begin{lstlisting}[basicstyle=\ttfamily]
/*(Before)*/
function withdrawBal () public{
  Balances[msg.sender] = 0;
  if(!msg.sender.send(Balances[msg.sender]))
    { revert(); }}
/*(After injection)*/
function withdrawBal () public{
  Balances[msg.sender] = 0;
  if(!msg.sender.send(Balances[msg.sender]))
    { //revert();
    }}
\end{lstlisting}
\caption{Weakening security example.}
\label{fig:weakening-example}
\end{figure}

\pdfoutput=1

\section{\sysname Algorithm} 
\label{sec:algorithm}

The process for injecting bugs takes as input the Abstract Syntax Tree (AST) of the smart contract, and has the following steps.
\begin{enumerate}
    \item Identify the potential locations for injecting bugs and generate an annotated AST marking all identified locations.
    \item Inject bugs into all marked locations to generate the buggy contract.
    \item Check the buggy contract using the evaluated tools and inspect the results for undetected bugs and false alarms.
\end{enumerate}
We discuss the steps in detail below.




\textbf{Bug locations identification:}
The AST is passed to \textit{Bug Locations Identifier}, that drives a bug injection profile (BIP) of all possible injection locations in the target contract for a given security bug. 
The BIP is derived using AST-based analysis for identifying potential injection locations in smart contract code by Algorithm \ref{alg:findlocs}. Algorithm \ref{alg:findlocs} takes as input the AST and the bug type to be injected, and outputs the BIP.

\begin{algorithm}[h]
\caption{Identifying Injection Locations Algorithm}
\label{alg:findlocs}
\small
\begin{algorithmic}[1]
\Procedure{FindAllPotentialLocations}{AST, bugType}
\For {Each form of code snippets in bugType}
    \If{snippetForm  == simple statement}
        \State $BIP \gets WalkAST(simpleStatement)$
    \ElsIf{snippetForm  == non-function block}
        \State $BIP \gets WalkAST(nonFunctionBlock)$
    \ElsIf {snippetForm  == functionDefinition}
        \State $BIP \gets WalkAST(functionDefinition)$
  \EndIf
 \EndFor
  \State $BIP \gets Find Related Security Mechanisms$
  \State $BIP \gets Find Code That Can Be Transformed$
 \State{return $BIP$}
 \EndProcedure
\end{algorithmic}
\end{algorithm}
\normalsize

To address the first challenge of identifying bug injection locations as mentioned in Section \ref{injectionchallenges}, we define rules that specify the relation between the bug to be injected and the target contract structure. In general, bugs take two forms: an individual statement, and a block of statements. A block of statements can be defined either as a stand-alone function, or a non-function block such as an 'if' statement. Therefore, we use a rule for each form of the bug that defines the specifications of the locations for injecting it. 

To identify such locations, for each distinct form of the code snippets defining the bug type to be injected, we walk the AST based on the code snippet form and the related rule (lines 2-10 in Algorithm \ref{alg:findlocs}). $WalkAST(simpleStatement)$, for example, will visit (parse) the AST and find all the locations where a simple statement can be injected without invalidating the compilation state of the contract, and the same for the other forms of the code snippets of the bug type. After identifying the locations for injecting code snippets of bugs, we also look for existing security mechanisms to be weakened to introduce the related bug, and the code patterns to be transformed for introducing the bug (lines 11 and 12). 

\textbf{Bug injection and code transformation:}
\sysname uses a systematic approach to inject bugs into the potential locations in the target contract. 
The \textit{Bug Injector} model seeds a bug for each location specified in the BIP. It uses text-based code transformation to modify the code where the information derived from the AST is used to modify the code to inject bugs.
Three different approaches are used to inject bugs as discussed in Section \ref{buginstrument}. In addition to injecting bugs in the target contract, \textit{Bug Injector} generates a \textit {BugLog} that specifies a unique identifier for each injected bug, and the corresponding location(s) in the target contract where it has been injected. 

\textbf{Buggy code check and results inspection:}
The resulting buggy contract is passed to the \textit{Tool Evaluator} that checks the buggy code using the tools under evaluation. 
It then scans the results generated by the tools looking for the bugs that were injected but undetected, with the help of \textit{BugLog} generated by the \textit{Bug Injector}. 
\sysname only considers the injected bugs that are undetected. So if an evaluated tool reported bugs in locations other than where bugs have been injected, \sysname does not consider them in its output of false negatives. 
This is to avoid potential vulnerabilities in the original contract from being reported by \sysname, which would skew the results. Moreover, \sysname inspects results generated by the tools looking for other reported bugs and checks if they are true bugs or false alarms (more details in Section~\ref{sec:eva}).

\section{Implementation}
\label{sec:imp}
\sysname approach is fully-automated (except for the pre-prepared buggy snippets). This involves compiling the code, injecting and generating buggy contracts, running the evaluated tools on the buggy contracts, and inspecting reports of the evaluated tools for false-negatives, mis-identified cases, and false-positives (except for the manual validation of the filtered false-positives).
To make \sysname reusable, we did not hard-code the patterns that are replaced to introduce bugs, but rather made them configurable from an external file. We have made {\em \sysname code publicly available}\footnote{
		\url{https://github.com/DependableSystemsLab/SolidiFI} }.

\sysname uses the Solidity compiler \textbf{solc} (supports compiler versions up to 0.5.12) to compile the source code of the smart contract to make sure it is free from compilation errors before bugs are injected. In addition, \sysname uses \textbf{solc} to generate the AST of the original code in JavaScript Object Notation (JSON) format. We have implemented the other components of \sysname in Python in about $1500$ lines of code. These components are responsible for identifying the potential locations for injection, injecting bugs using a suitable approach, generating the buggy contract, and inspecting the results of the evaluated tools and then reporting the undetected bugs and false alarms. Finally, we developed a Python client to interact with contracts deployed on Ethereum network - this client is used for assessing the exploitability of the injected bugs in the generated buggy contracts. 

\pdfoutput=1

\section{Evaluation} 
\label{sec:eva}
The aim of our evaluation is to measure the efficacy of \sysname in evaluating smart contract static analysis tools, and finding cases of undetected bugs (i.e., false negatives) and false positives. We also measure the performance of \sysname itself, as well as the ability to exploit the undetected bugs. We made all the experimental artifacts used in this study and our results publicly available\footnote{ \url{https://github.com/DependableSystemsLab/SolidiFI-benchmark}}. Our evaluation experiments are thus derived to answer the following research questions:

\begin{itemize}
\item {\textbf{ RQ1.} \it What are the false negatives of the tools being evaluated?}
\item {\textbf{ RQ2.} \it What are the false positives of the tools being evaluated?}
\item {\textbf{RQ3.} \it Can the injected bugs in the contracts be activated (i.e., exploited) at runtime by an external attacker?}
\item {\textbf{RQ4.} \it What is the performance of \sysname?}
\end{itemize}


As mentioned earlier, we have selected six static analysis tools for evaluation, Oyente \cite{luu2016making}, Securify \cite{tsankov2018securify}, SmartCheck \cite{tikhomirov2018smartcheck}, Mythril \cite{mueller2018smashing}, Manticore \cite{mossberg2019manticore}, and Slither \cite{feist2019slither}. We downloaded these tools from their respective online repositories, which are mentioned in the corresponding papers (Oyente 0.2.7, Securify v1.0 as downloaded and installed in Dec 2019, Mythril 0.21.20,  Smartcheck 2.0, Manticore 0.3.2.1, and Slither 0.6.9).

To perform our experiments, we used a data set of fifty smart contracts, chosen from the list of verified smart contracts available on \textit{Etherscan} \cite{etherscan}, a public repository of smart contracts written in Solidity for Ethereum. We selected these contracts based on three factors namely (1) code size (we selected contracts with different sizes that were representative of Etherscan contracts ranging from small contracts with tens of lines of code to large contracts with hundreds of lines of code), (2) compatibility with Solidity version 0.5.12 (at the time of writing, {\em 312 out of 500 verified smart contracts} in EtherScan supported Solidity 0.5x and higher), and (3) contracts with a wide range of functionality (e.g., tokens, wallets, games). 
Table \ref{tab:benchmark} shows the number of lines of code (including comments), and number of functions and function modifiers\footnote{Function modifier checks a condition before the execution of the function.} for each contract. The contracts range from 39 to 741 lines of code (loc), with an average of 242 loc.

 We limited ourselves to 50 contracts due to the time and effort needed to analyze the contracts by the evaluated tools and inspect the analysis results of the tools to verify false-positives. {\em With that said, even with this dataset, SolidiFI found significant numbers of undetected bugs in the tools (e.g., false negatives), as will be discussed in the following sections.} 

As explained in Section \ref{sec:SolidiFI}, in our experiments, we injected bugs belonging to seven different bug types within the detection scope of the selected tools. Table \ref{tab:bugsandtools} shows the bug types, and the tools that are designed to detect each bug type. We chose these bug types based on the bug types detected by the individual tools, and because these bugs are common in smart contracts, and lead to vulnerabilities that have been exploited in practice~\cite{dao2017,batchOverflow, EthereumAttacksHistory}. However, \sysname is not confined to these bug types. 

In our experiments, we set the time-out value for each tool to 15 minutes per smart contract and bug type. If a tool's execution exceeds this timeout value, 
we terminate it and consider the bugs found as its output. While 15 minutes may seem high, our goal is to give each tool as much leeway as possible.
Only 2 of the tools exceeded this time limit in some cases (i.e., Mythril and Manticore). For these two tools, we experimented with larger timeout values, 
but they did not significantly increase their detection coverage. Note that the total time taken to run the experiments was already quite high with this
timeout value - for example, it took us about 4 days to analyze the contracts using Mythril (50-contract*6 bug-types*15-minute = 75 hours). 

\begin{table}[h]
\begin{center}
\caption{Contracts benchmark. F represents Functions, and M represents Function Modifiers}
\begin{tabular}{|c|c|c||c|c|c||c|c|c|}  
\hline
 \textbf{Id}&{\rotatebox{90}{\textbf{Lines}}}&{\rotatebox{90}{\textbf{F+M}}}& \textbf{Id}&{\rotatebox{90}{\textbf{Lines}}}&{\rotatebox{90}{\textbf{F+M}}}& \textbf{Id}&{\rotatebox{90}{\textbf{Lines}}}&{\rotatebox{90}{\textbf{F+M}}} \\
\hline
1&103&6&18&406&29&35&317&29\\
\hline
2&128&9&19&218&32&36&383&20\\
\hline
3&132&10&20&308&27&37&368&24\\
\hline
4&117&6&21&353&18&38&195&24\\
\hline
5&250&17&22&383&19&39&52&4\\
\hline
6&161&22&23&308&20&40&465&22\\
\hline
7&165&22&24&741&27&41&160&8\\
\hline
8&251&17&25&196&12&42&128&16\\
\hline
9&249&19&26&143&20&43&285&22\\
\hline
10&39&5&27&336&33&44&298&24\\
\hline
11&193&19&28&195&24&45&156&14\\
\hline
12&281&27&29&312&13&46&125&6\\
\hline
13&161&8&30&711&57&47&223&18\\
\hline
14&185&20&31&216&12&48&232&19\\
\hline
15&160&8&32&143&14&49&52&4\\
\hline
16&248&27&33&129&16&50&171&18\\
\hline
17&128&17&34&445&29&&&\\
\hline
\multicolumn{7}{|c|}{\textbf{Average values }}&\textbf{242}&\textbf{18}\\
\hline
\end{tabular}
\label{tab:benchmark}
\end{center}
\end{table}

\begin{table}[h]
\begin{center}
\caption{Bug types used in our evaluation experiments: '*' means that the tool can detect the bug type.}
\begin{tabular}{|c|c|c|c|c|c|c|}  
\hline
 \textbf{Bug Type}&{\rotatebox{90}{\textbf{Oyente}}}&{\rotatebox{90}{\textbf{Securify}}}&{\rotatebox{90}{\textbf{Mythril}}}&{\rotatebox{90}{\textbf{SmartCheck}}}&{\rotatebox{90}{\textbf{Manticore}}}&{\rotatebox{90}{\textbf{Slither}}} \\
 \hline
 Re-entrancy &* & * & * & *&* &*\\
\hline
Timestamp dependency &* & & * & *&&* \\
\hline
Unchecked send & & * & * & &&\\
\hline
Unhandled exceptions &* & * & * & *&&* \\
\hline
TOD&* & *& & &&\\
\hline
Integer overflow/underflow & *& & * & *&* &\\
\hline
Use of tx.origin & & & * & * &&*\\
\hline
\end{tabular}
\label{tab:bugsandtools}
\end{center}
\end{table}

\smallskip \subsection{RQ1:What are the false negatives of the tools being evaluated?}

The core part of our evaluation is to use \sysname to inject bugs, and evaluate the effectiveness of the tools in detecting the injected bugs. 
We performed the following steps in our experiments. First, \sysname is used to inject bugs of each bug type in the code of the 50 smart contracts, one bug type at a time. 
The resulting buggy contracts are then checked using the static analysis tools. 
Finally, 
the number of the injected bugs that were not detected by each tool were recorded. 


To get meaningful results, we inject bugs that are as distinct as possible by preparing diverse set of distinct code snippets with different data inputs and function calls- this resulted in $9369$ distinct bugs.
We consider two injected bugs as distinct if the static analysis tool under study would reason about them differently based on the underlying methodology, where either the data and control flow leading to the injected bug is different, or the design patterns of the bug snippets are different. 
{\em To  ensure a fair evaluation, we inject only the bugs that are supposed to be detected by each tool. }

We consider an injected bug as being correctly detected by a tool if and only if it identified both the line of code in which the bug was injected, as well as the bug type (e.g., Re-entrancy). 
In many cases, we observed that the tool would correctly identify the line of code in which the bug occurred, but would misidentify the bug type. Therefore,
we also report the former separately.

The results of injecting bugs of each bug type, and testing them using the six tools are summarized in Table \ref{tab:flasenegative}. 
 In the table, \enquote{Injected bugs} column specifies the total number of injected bug for each bug type, \enquote{\checkmark}  means we did not find any undetected bug of that bug type (row), while \enquote{NA} means the bug type is out of scope of the tool, i.e., it is not designed to detect the bug type. The numbers for each column specify the total number of bugs that were either incorrectly detected or not detected by the tool corresponding to the bug type specified in that row.
 The number within parentheses specifies the number of cases that were not reported by the tool - this does not consider the incorrect reporting of the bug type.
 
 
 From the table, we can see that a significant number of false negatives occur for all the evaluated tools,
 and that {\em none of the tools was able to detect all the injected bugs correctly even if we accepted a incorrect bug type with the correct line number as a detected bug}.
 In fact, the only tool that had 100\% coverage for individual bug types was Slither, for {\em Reentrancy} and {\em tx.origin} bugs.
 Of all tools, Slither had the lowest false-negatives, followed by Securify across bug types.

Our results thus show that all static analysis tools have many corner cases of bugs that they are not able to detect.
{\em Note that it is surprising that our technique found as many undetected bugs by the tools as it did, given that our goal was not specifically to exercise corner cases of the tools in question.} 
We will discuss the reasons for the missed detections and the implications later (Section~\ref{sec:discussion}).

\begin{table}[h]\footnotesize
\begin{center}
\caption{False negatives for each tool. Numbers within parantheses are bugs with incorrect line numbers or unreported.}
\begin{tabular}{|c|c|c|c|c|c|c|c|}
\hline
 \textbf{Security bug} &{\rotatebox{90}{\textbf{Injected bugs}}}&{\rotatebox{90}{\textbf{Oyente}}} &{\rotatebox{90}{\textbf{Securify}}}&{\rotatebox{90}{\textbf{Mythril}}}&{\rotatebox{90}{\textbf{SmartCheck}}}&{\rotatebox{90}{\textbf{Manticore}}}&
 {\rotatebox{90}{\textbf{Slither}}}\\
 \hline
 Re-entrancy&1343 &\shortstack{1008 \\ (844)}&\shortstack{232\\(232)}&\shortstack{1085\\(805)}& \shortstack{1343\\(106)}&\shortstack{1250\\(1108)}&\cellcolor{asparagus}\checkmark\\
\hline
Timestamp dep&1381 &\shortstack{1381\\(886)}&\cellcolor{lightgray}NA&\shortstack{810\\(810)}& \shortstack{902\\(341)}&\cellcolor{lightgray}NA&\shortstack{537\\(1)} \\
\hline
Unchecked-send&1266 &\cellcolor{lightgray}NA&\shortstack{499\\(449)}& \shortstack{389\\(389)} & \cellcolor{lightgray}NA&\cellcolor{lightgray}NA &\cellcolor{lightgray}NA\\
\hline
Unhandled exp&1374 &\shortstack{1052\\(918)}
&\shortstack{673\\(571)}& \shortstack{756\\(756)} &\shortstack{1325\\(1170)}&NA&\shortstack{457\\(128)}\\
\hline
TOD &1336& \shortstack{1199\\(1199)} &\shortstack{263\\(263)}&\cellcolor{lightgray}NA&\cellcolor{lightgray}NA&\cellcolor{lightgray}NA&\cellcolor{lightgray}NA\\
\hline
Integer overflow&1333 &\shortstack{898\\(898)}&\cellcolor{lightgray}NA&\shortstack{1069\\(932)}&\shortstack{1072\\(1072)}&\shortstack{1196\\(1127)}&\cellcolor{lightgray}NA\\
\hline
tx.origin &1336&\cellcolor{lightgray}NA&\cellcolor{lightgray}NA&\shortstack{445\\(445)}&\shortstack{1239\\(1120)}&\cellcolor{lightgray}NA&\cellcolor{asparagus}\checkmark\\
\hline
\end{tabular}
\label{tab:flasenegative}
\end{center}
\end{table}

 \subsection{RQ2: What are the false positives of the tools being evaluated?}
 
A false-positive occurs when a tool reports a bug, but there was no bug in reality. Unlike false negatives, where we know exactly where the bugs have been injected, and hence have ground truth, measuring false positives is challenging due to the lack of ground truth. 
This is because we cannot assume that the smart contracts used are free of bugs (though they are chosen from the verified contracts on Etherscan). 
Further, manually inspecting each bug report and related contract involves a tremendous amount of effort due to the large number of bug reports, and is hence not practical.

To keep the problem of determining false-positives tractable, we came up with the following approach.
The main idea is to manually examine only those bugs that are {\em not} reported by the majority of the other tools for each smart contract. In other words, we
conservatively assume that a bug that is reported by a majority of the tools cannot be a false positive. However, at worst, we will underestimate
the number of false positives in this approach, subject to the vagaries of the manual inspection process. We also verified that many of the bugs
that are excluded by the majority are indeed false-positives by manually examining a random sample of them.
 
Even after this filtering, we had to manually inspect a significant
number of bugs to determine if they were false positives. Therefore, for each tool, we randomly selected 20 bugs of each bug type category that were not excluded by the majority approach, and inspected them manually. For those cases where the number of bugs is less than or equal to 20, we inspected them all. 
Based on the results of our manual inspection, we estimated the false positives as the percentage of bugs inspected that were indeed false positives, multiplied by the number of bugs filtered (i.e., not excluded).

For example, assume that the total number of bugs reported by a tool is $100$. Of these 100 bugs, let us assume that $60$ are also reported by the majority of the other tools for the smart contract, and hence we exclude them. Of the remaining $40$ filtered bugs, we manually examine $20$ bugs chosen at random. Assume that $16$ of these are indeed false-positives. 
We assume that 80\% of the filtered bugs are false-positives, and estimate the number of false-positives to be $32$. 

 
The results of false positives reported by each tool are summarized in Table \ref{tab:flasepositive}. In the table, the \enquote{Threshold} column refers to the majority threshold, which is the number of tools that must detect the bug in order for it to be excluded from consideration - this number depends on how many tools are able to detect the bug type. 
For each tool, the sub-column \enquote{Reported} shows the number of bugs reported by the tool, 
the sub column \enquote{FIL} shows the number of bugs that have been filtered (not excluded) by the majority approach,
 while the sub column \enquote{FP} shows the false positives of the tool based on the manual inspection as explained above. 
 Empty cells in the table represent cases where a tool was not designed to detect a bug type. 
 Note that some of the tools detected bugs outside the 7 categories that we considered - we called these as {\em miscellaneous}.

From the table, we can see that all the evaluated tools have reported a number of false positives, ranging from $2$ to $801$ for most of the bug types.
Interestingly, the results show that the tools with low numbers of false negatives reported high false positives, i.e., {\em Slither} and {\em Securify}. 
For example, although {\em Slither} was the only tool that successfully detected all the injected Re-entrancy bugs, it reported significant false positives.
This raises the question of whether the high detection rate was simply a result of over-zealously reporting bugs by the tool (this is
also borne out by the high number of bugs reported under the {\em miscellaneous} category by this tool).
{\em  This highlights the need for security analysis tools that are able to detect bugs while maintaining low false positive rates.}
 
We provide some examples of the false-positive cases below. 
For example, most unhandled exception bugs were reported even though the code checks the return values of the send functions for exceptions using {\em require()}. As another example, many false positives were re-entrancy bugs where the code contains the required checks of the contract balance, and updates the contract states before the Ether transfer. Oyente reported several cases as integer over/underflow even though they are no integer related calculations (e.g., {\em string public symbol = "CRE";}). On the other side, we tried to be consistent with the assumptions considered by the tools during our manual inspection. For example, some of the cases that we considered as true bugs were re-entrancy bugs  that use the {\em transfer} function. This function protects against re-entrancy issues as it has limited gas; however, we considered them as true bugs as the attack can happen if the gas price changes - this is detected by some of the tools (e.g., Slither).

\begin{table*}[h]\footnotesize
\begin{center}
\caption{False positives reported by each tool. Empty cells mean that the tool was not designed for that particular bug type.}
\begin{tabular}{|c|c|c|c|c|c|c|c|c|c|c|c|c|c|c|c|c|c|c|c|c|c}  
\hline
 \textbf{Bug Type}&{\rotatebox{90}{\textbf{Threshold}}}&\multicolumn{3}{|c|}{\rotatebox{90}{\textbf{Oyente}}}&\multicolumn{3}{|c|}{\rotatebox{90}{\textbf{Securify}}}&\multicolumn{3}{|c|}{\rotatebox{90}{\textbf{Mythril}}}&\multicolumn{3}{|c|}{\rotatebox{90}{\textbf{SmartCheck}}}&\multicolumn{3}{|c|}{\rotatebox{90}{\textbf{Manticore}}}&\multicolumn{3}{|c|}{\rotatebox{90}{\textbf{Slither}}} \\
 \hline
= &&Reported&FIL&FP&Reported&FIL&FP&Reported&FIL&FP& Reported&FIL&FP&Reported&FIL&FP&Reported&FIL&FP\\
 \hline
 Re-entrancy & 4& 0 &0&- &12 & 12&12&54&54&43&0&0&-&6&6&6&79&79&71\\
\hline
Timestamp dep & 3& 0&0 &- & &&& 12&12&0&0&0&-&&&&12&12&0\\
\hline
Unchecked send & 2& &  &&7 &4&4&14&3&3&&&&&&&&&\\
\hline
Unhandled exp & 3&10  &10 & 10 &0&0&-&0&0&-&6&6&6&&&&0&0&- \\
\hline
TOD&2 &32 &24& 24 &121 &97&97&&&&&&&&&&&&\\
\hline
Over/under flow &3 &947 &943 &801 & && &17&3&3&3&2&2&9&9&9&&&\\
\hline
Use of tx.origin &2 & &&  & &&&0&0&-&3&1&0&&&&4&2&0\\
\hline
Miscellaneous && 0  &&&318&&&144&&&1520&&&169&&&1807&&\\
\hline
\end{tabular}
\label{tab:flasepositive}
\end{center}
\end{table*}

\smallskip \subsection{RQ3: Can the injected bugs be activated by an external attacker?}
The goal of this RQ is to assess whether the undetected bugs in RQ1 can be activated in the contract at runtime. This is to determine whether the reason behind the bug not being detected by the evaluated tool was because the bug cannot be activated (and hence cannot be exploited by an attacker). We deploy the set of buggy contracts with the undetected bugs (found in RQ1) on the Ethereum blockchain, and execute transactions that attempt to activate them. 

To conduct these experiments, we use MetaMask \cite{MetaMask}, a browser extension that allows us to connect to an Ethereum node called INFURA \cite{INFURA}, and run our buggy contracts on this node. We have created Ethereum accounts on Ethereum Kovan Testnet (test network) using MetaMask, and deposited sufficient amount of Ether to these accounts to enable us to execute transactions (pay the required gas for transactions). We use Remix \cite{remix} (Solidity editor) to deploy contracts on Ethereum Kovan Testnet.  Remix enables us to connect with MetaMask to deploy contracts on INFURA Ethereum node. 


We illustrate the process with an example. As mentioned in RQ1, Manticore did not report instances of injected integer overflow/underflow bugs - an example is shown in Figure \ref{fig:undetected-underflow}. Our goal is to attempt to activate this bug in the deployed buggy contract by calling the function \textit{bug\_intou3()}. The returned result was 246 - this is not the expected value (-10) due to the use of unsigned integer type (i.e., uint8) instead of a signed integer type, which resulted in an integer underflow. Thus, the bug can be activated by an attacker. 

We had to manually craft inputs for each bug in order to test its activation, which takes significant effort. 
 Because of the large number of undetected bugs, we selected 5 undetected bugs for each bug type randomly from different contracts to test their activation. Table \ref{tab:bugsactivity} shows the results of our activation experiments.  In the table \enquote{--} means we were not able to perform experiments on this bug type, as it requires the attacker to behave as a miner, which would consume a significant amount of computational resources. The results show that one can exploit (activate) all the selected bugs in their related buggy contracts. 
 Therefore, the infeasibility of activation of the bug was not the reason that the evaluated tools failed to detect the injected bugs. 

\begin{figure}[h]
\footnotesize
\centering
\begin{lstlisting}[basicstyle=\ttfamily]
function bug_intou3() public{
    uint8 vundflw =0;
    vundflw = vundflw -10; // underflow
    return vundflw;}
\end{lstlisting}
\caption{Undetected integer underflow bug}
\label{fig:undetected-underflow}
\end{figure}


\begin{table}[h]
\begin{center}
\caption{Activity of Selected Undetected Bugs. }
\begin{tabular}{|c|c|c|}  
\hline
 \textbf{Bug type}&\textbf{Selected bugs}&\textbf{Activated bugs }\\
 \hline
 Re-entrancy &5 &5  \\
\hline
Timestamp dependency &5 & 5\\
\hline
Unchecked send &5 &5 \\
\hline
Unhandled exceptions &5 &5 \\
\hline
TOD& -- & -- \\
\hline
Integer overflow/underflow &5 &5 \\
\hline
Use of tx.origin &5 &5 \\
\hline
\end{tabular}
\label{tab:bugsactivity}
\end{center}
\end{table}

\smallskip \subsection{RQ4: What is the performance of \sysname?} 
Finally, we measured the performance of \sysname in terms of the time it takes to inject bugs and generate buggy contracts. We excluded the time of running the tools being evaluated to check the buggy contracts, as this is tool-specific and independent of \sysname. 
We preformed injection of each bug type in each contract five times and  calculated the average of the five runs, and then calculated the average of injecting the seven bug types in each contract. The average time of injecting all instances of bug types in a contract is 25 seconds, and the worst case time was 46 seconds (for contract 24, which was the largest contract in our set). {\em Thus, \sysname takes less than 1 minute on average per contract and bug type. } 

\pdfoutput=1

\section{Discussion}
\label{sec:discussion}

In this section, we examine the reasons for the false negatives of the tools observed in RQ1.
We then examine the implications of the results, and our methodology, on both tool developers and end users.
Finally, we examine some of the limitations of \sysname and threats to validity of our experiments. 

\subsection{Reasons for False-Negatives}

To establish a practical understanding of the presented results and why some bugs were not detected, we will highlight the code snippets for some of the bugs that were not detected, and then discuss the reasons behind them. We organize this discussion by tool.

\textbf{Oyente} was not able to detect many instances of injected re-entrnacy, timestamp dependency, unhandled exceptions, integer overflow/underflow, and TOD bugs as mentioned earlier
\footnote{Because the released version of Oyente did not work with the latest version of the Solidity compiler (0.5.12), we made few changes to the injected contracts to get it to work with Oyente - these did not impact the tool's coverage.}.
According to the paper \cite{luu2016making}, {\em Oyente} works on detecting only re-entrancy bugs that are based on the use of {\em call.value}. Some of the recent tools, such as {\em Slither}, consider the detection of re-entrancy bugs with limited gas that are based on {\em send} and {\em transfer}. Those papers claim that {\em send} and {\em transfer} do not protect from re-entrancy bugs in case of gas price change. Furthermore, {\em Oyente} failed to detect instances of re-entrancy bugs that are based on the use of {\em call.value}.
One of the TOD code snippets we used in our experiments is mentioned in the running example at Figure~\ref{fig:example-contract} on lines 9-16, which emulates a simple game and its winner.  The malicious behavior occurs when the two transactions are executed in one block and the attacker tries to change the order of the received transactions. To understand why this bug is not detected by Oyente, in Oyente the EVM bytecode is represented as a control flow graph (CFG). The execution jumps are used as edges that connect the nodes of the graph representing the basic execution blocks in the code. The symbolic execution engine of Oyente uses the CFG to produce a set of symbolic traces (execution paths), each associated with a path constraint and auxiliary data, to verify pre-defined properties (security bugs being detected). Basically, Oyente detects TOD by comparing the different execution paths and the corresponding data flow (Ether flow) for each path. Oyente reports those different execution paths that have different Ether flows. 

For Oyente to be able to detect all TOD bugs successfully, the symbolic execution engine should generate all possible execution paths for the contract to find the erroneous path  - this is challenging due to the incompleteness of symbolic execution. 
It also uses some bounds that limit the symbolic execution.

{\bf Smartcheck} failed to detect most of the injected bugs across all the categories. Smartcheck checks for bugs by constructing an Intermediate Representation (IR) from the source code, and then using XPath patterns to search for bugs in the IR. This approach lacks accuracy as some bugs cannot be expressed as XPath expressions. For example, the re-entracy bug is difficult to express as an XPath pattern, and is hence not detected.


Further, because Smartcheck uses XPath patterns that detect specific syntax of some bugs,  even a slight variation in the syntax of the bug snippets would not match the XPath patterns. For instance, SmartCheck did not report some occurrences of unhandled exceptions.
By checking the code snippet for one of the undetected unhandled exception bug depicted in Figure~\ref{fig:mishandledexp1},
we found that Smartcheck was not able to detect it as unchecked send because Smartcheck only looks for {\em send} functions without an if-statement (that checks the return value). However, in this snippet, the send is within an if-statement even though the {\em revert()} exists in the else clause of the if-statement, which will be triggered on the successes of send. The same happens when other functions are used for sending ether  (e.g., call, etc.) instead of {\em send}. This is an inherent problem with syntactic, rule-based tools such as Smartcheck.

 \begin{figure}[h]
\centering
\footnotesize
\begin{lstlisting}[language= Solidity, basicstyle=\ttfamily]
if (!addr.send (42 ether))
     {receivers +=1;}
else
     {revert();}}
\end{lstlisting}
\caption{Unhandled exception code snippet 1.}
\label{fig:mishandledexp1}
\end{figure}

{\bf Mythril} was the tool with the largest set of undetected bugs in our experiments. It failed to detect many instances of re-entrancy, timestamp dependency, unchecked send, unhandled exceptions, integer overflow/underflow  and use of tx.origin. For example, the buggy code in Figure~\ref{fig:mishandledexp2} was not detected by Mythril. The condition of the if-statement that checks the return value of the send will always be evaluated as true because of the added condition \enquote{$||$ 1==1}. Hence, the execution of the contract would be reverted in all cases by the function {\em revert()}, regardless of whether the send succeeds. {\em This is incorrect as the execution should only be reverted on the fails of send.} However, Mythril does not detect this as it only evaluates the {\em send()} part in the condition of the if-statement rather than evaluating the whole condition with the OR part ($||$ 1==1).

{\em Mythril} is also very slow in term of the time it takes to analyze contracts. 
Although we set the time-out for analyzing each contract to 15 minutes, as mentioned earlier, we also tried setting the timeout to 30 minutes and did not observe 
any increase in the number of bugs demonstrating that increasing the time-out has diminishing returns. 
We also found the number of undetected bugs increase in the large contracts, as Mythril enumerates symbolic traces and this does not scale well in large contracts. 

\begin{figure}[h]
\centering
\footnotesize
\begin{lstlisting}[language= Solidity, basicstyle=\ttfamily]
if (!addr.send (10 ether) || 1==1)
     {revert();}
\end{lstlisting}
\caption{Unhandled exception code snippet 2.}
\label{fig:mishandledexp2}
\end{figure}

Like the other tools, Mythril also misreported the types of many of the injected bugs. 
Figure~\ref{fig:buggy_code} shows part of a buggy contract injected using \sysname. The injected contract allows users to manage their tokens and send tokens to each other. We injected a re-entrancy bug using \sysname in the contract at lines 185-188. However, Mythril reported the re-entrancy bug as  "Unchecked Call Return Value" (i.e., Unhandled exception) at line 186. By inspecting this line of code, we can see that the return value of the {\em send} function is checked and the balance is reset to zero on the success of {\em send}, so there is no unhandled exception as reported. 
This calls into question Mythril's soundness in detecting this type of bugs as well as its completeness in detecting Re-entrancy bugs.

\begin{figure}[h]
\centering
\footnotesize
\begin{lstlisting}[language= Solidity, basicstyle=\ttfamily, firstnumber=177]
function transfer(address _to, uint256 _value) public returns (bool success) {
    require(balances[msg.sender] >= _value);
    balances[msg.sender] -= _value;
    balances[_to] += _value;
    emit Transfer(msg.sender, _to, _value); 
    return true;
    }
    
function withdraw_balances_re_ent36 () public {
   if (msg.sender.send(balances[msg.sender ]))
        balances[msg.sender] = 0;
      }
\end{lstlisting}
\caption{Part of a buggy contract injected by reentrancy bug.}
\label{fig:buggy_code}
\end{figure}



{\bf Manticore} was not able to detect instances of re-entrancy and integer overflow/underflow. Unlike other evaluated tools employing symbolic executions, we noticed that Manticore takes a long time to analyze smart contracts, and in some cases it times-out. 
It consumes significant memory space as well on our system. 
Moreover, Manticore crashed and failed to analyze most of the contracts and threw exception errors. The 50 main contracts  used in our experiments consist of 123 analyzable contracts (each contract file may contain more than one contract). Out of them, {\em Manticore} crashed for 83 contracts injected by re-entrancy bugs and 73 contracts injected by integer overflow bugs. 
 We reached out to the tool developers to get fixes or explanation for these issues; however, there was no response (as of the time of submission). 

{\bf Securify} was not able to detect several cases of injected bugs belonging to re-entrancy, unchecked send, unhandled exceptions, and TOD. In addition, we found many cases where {\em Securify} failed to analyze the injected contracts and threw an error.
Out of the 200 contracts injected by the four bug types (50 contracts for each bug type), {\em Securify} failed to analyze 5 contracts injected by unhandled exceptions, 4 injected by re-entrancy, 4 injected by unchecked send, and 4 injected by TOD bugs. If we excluded the injected bugs in those contracts, the number of undetected bugs by {\em Securify} will be as following: (re-entrancy: 105, unchecked send: 332, unhandled exceptions: 402, and TOD: 136). 
{\em Securify} also reported a high number of TOD false positives compared with {\em Oyente}.  A recent study \cite{Feng2019smartscopy} found that the reported false alarms by Securify are due to over-approximation of the execution. 

{\bf Slither} Although {\em Slither} has almost 100\% accuracy in detecting re-entrancy, timestamp, and tx.origin bugs, it was not able  to detect many instances of unhandled exceptions. Moreover, it had high number of re-entrancy false positives as mentioned earlier. 

\subsection{Implications}
\textbf{Tool Developers}
There are two implications for tool developers. The first implication is that using pattern matching for detecting bugs, especially by employing simple approaches such as XPaths matching, is not an effective way for detecting smart contract bugs for the reasons mentioned earlier in this section. The second point is that bug detection approaches that are based on enumerating symbolic traces are impeded by path explosion and scalability issues. 
Therefore, there is a need for sophisticated analysis tools that also consider the semantics of the analyzed code instead of depending only on analyzing the syntax and symbolic traces. For example, static analysis might work better if combined with formal methods that consider the semantic specifications of Solidity and EVM. Recent papers \cite{grishchenko2018semantic, hildenbrandt2017kevm, hirai2017defining, bhargavan2016formal, amani2018towards} have proposed semi-automated formal verification for performing analysis of smart contracts.

\textbf{End Users of Tools}: For smart contract developers, who are the end users of the static analysis tools, there are three implications.
First, they can use \sysname to assess the efficacy of static analysis tools to choose the most reliable tools with no or low false negatives for their use cases. Second, developers should not rely exclusively on static analysis tools, and should test the developed contracts extensively. \sysname can help them build test suites by introducing mutations and checking if the test cases can catch them. Finally, the generated bugs by \sysname and their relative locations in the code can be used for educating developers on writing secure code.

\subsection{Limitations of \sysname}

There are two limitations of \sysname. First, the current version of \sysname works only on Solidity static analysis tools. Although Solidity is the most common language for writing Ethereum smart contracts and most of the proposed tools target analysis of Solidity contracts, \sysname functionality can be easily extended to other languages. 
Secondly, the bug injection approach employed by \sysname requires pre-prepared code snippets (for each bug type), which requires some manual effort. However, this is a one-time cost for each bug type (we have provided these as part of the tool).  
 
 \subsection{Threats to Validity}
 
 An external threat to the validity is the limited number of smart contracts considered, namely 50. We have mitigated this threat by considering a wide-range of smart contracts with varying functionality and code sizes.  We emphasize that SolidiFI covers all syntactic elements of Solidity up to version 0.5.12 and, our data-set contains a wide variety of contracts with different features (e.g., loops). Also, we  selected contracts with different sizes that were representative of EtherScan contracts ranging from 39 to 741 locs, with an average of 242 loc (Table~\ref{tab:benchmark}).  
 
 There are two internal threats to validity. First, the number of tools considered is limited to 6. However, as mentioned, these represent the common tools used in other studies on smart contract static analysis. Further, they are widely used in both academia and industry. All the tools available are open-source and are being actively maintained (with the exception of Oyente). Further, the implemented prototype of \sysname is reusable and can be easily extended to evaluate other tools. The second internal threat to validity is that we only injected 7 bug types. However, these bug types have been (1) considered by most of the tools evaluated, and (2) exploited in the past by real attacks. Therefore, we believe they are representative of security bugs in smart contracts.
 
 Finally, a construct threat to validity is our measurement of false-negatives and false-positives. For false-negatives, it is possible that the bugs cannot be exploited in practice. We have partially mitigated this threat by sampling the set of false-negatives and attempting to exploit them (RQ3). For false-positives, it is possible that the reported bugs are true positives. Again, we have partially mitigated this threat by conservatively considering the bugs reported by the majority of the tools for each bug category as true positives. 

\pdfoutput=1

\section{Related Work}
\label{sec:relWork} 

\textbf{Bug-finding Tools Evaluation}
The approach of injecting bugs for evaluating the effectiveness of bug finders has been applied in other contexts than smart contracts.
Bonett et al. proposed $\mu$SE \cite{bonett2018discovering}, a mutation-based framework for the evaluation of Android static analysis tools that works as follows. First, a fixed set of  security operators are created describing the unwanted behavior that the tools being evaluated aim to detect (e.g., data leakage). 
Then, $\mu$SE inserts the security operators into mobile apps based on mutation schemes, that consider Android abstractions, tools reachability and security goals of the tools, thereby creating multiple mutants that represent unwanted behavior within the apps. 
The mutated apps are analyzed using the static tools to be evaluated. However, unlike our work, the undetected mutants are analyzed manually. Further, their framework focuses only on data leak detection tools, unlike our work, which is more general.

Pewny et al. \cite{pewny2016evilcoder} automatically find potential vulnerable  locations in C code, and modify the source code to make it vulnerable. Program analysis techniques are used to find sinks in the programs matching specific bug patterns, and find connections to user-controlled sources through data-flow. The program is modified accordingly to make it exploitable. In contrast to our approach, the vulnerable code locations to be injected are randomly chosen, and the implemented prototype targets the injection of spatial memory errors through the modification of security checks.

Dolan-Gavitt et al. \cite{dolan2016lava} proposed LAVA for generating and injecting bugs into the source code of programs using dynamic taint analysis. A guard is inserted for every injected bug for triggering the vulnerability if a specific value occurs in the input. Specifically, LAVA identifies an execution trace location where an unmodified and dead input data byte (DUA) is available. Then code is added at this location to make the DUA byte available, and use it execute the vulnerability.
Unlike our work, the injection is based on dynamic taint analysis, and the injected bugs are accompanied by triggering inputs. In contrast, our goal is to transform invulnerable code to systematically introduce vulnerabilities in it.

In recent work, Akca et al. \cite{akca2019solanalyser} proposed a tool that the authors used to compare the effectiveness of their introduced smart contracts static analyzer with some other tools. by injecting a single bug into the contract code. Unlike our approach that injects exploitable bugs into all potential locations in the contract code, the tool uses Fault Seeder \cite{peng2019sif} to generate contract mutants by injecting only a single bug snippet (hard-coded in the source code) into a specific location in the smart contract. In addition, the authors conducted manual inspection of the tool reports to determine false negatives. Injecting a single bug does not provide a comprehensive coverage evaluation of static analysis tools. Also, it is not clear how to evaluate the efficacy of static analysis tools on detecting deep vulnerabilities and corner cases by injecting only a single bug. As presented before, each bug can be introduced in the code in several ways, in this case, injecting a single-bug will not test the efficacy of the static analysis tools to detect various variants of each bug.

Durieux et al. \cite{durieux2019empirical} compared a number of smart contract static analysis tools. Unlike our work, the evaluation is based on using 69 manually annotated smart contracts with 112 bugs in total. The vulnerable contracts are collected from online repositories that are not agnostic of the evaluated tools, hence, results might be biased (e.g., collecting 50\% of the contracts from SWC Registry referenced by Mythril and maintained by the team behind it). In contrast, our goal is to perform systematic and comprehensive coverage evaluation of static analysis tools by transforming invulnerable code to systematically introduce vulnerabilities into all valid locations. We evaluated 6 tools on detecting about 9369 distinct bugs. To provide a fair evaluation of the tools, we evaluate each tool only on the bugs that it is designed to detect. Our proposed approach can be easily used to evaluate smart contract static analysis tools for detecting other bug types. Further, it enables end-users to choose any dataset of smart contracts for evaluating the tools.

\textbf{Smart Contract Testing and Exploitation}:
There have been many recent papers on testing smart contracts, and automatically generating security exploits on them.
Wu et al. \cite{wu2019mutation} produce test cases by mutating specific patterns in smart contracts.
Chan et al. \cite{chan2018fuse} develop a fuzz testing service (Fuse) to support the fuzz testing of smart contracts. This is a work in progress report, and only presents the architecture of the fuzz service being developed. Wang et al. \cite{wang2019towards} target the generation of test suites for smart contracts. This work guides automatic generation of test cases for Ethereum smart contracts.
Eth-mutants \cite{eth-mutants} is another mutation testing tool for smart contracts. However, it is limited to replacing $<$ to $\leq$, and $>$ to $\geq$ (and vice versa).
Unlike our focus on evaluating smart contracts' static analysis tools, these papers target the generation of test cases for the smart contracts.

Other papers focus on automatic exploitation of smart contracts to generate exploits or malicious inputs to exploit found code vulnerabilities \cite{krupp2018teether, jiang2018contractfuzzer, Feng2019smartscopy}.  teEther \cite{krupp2018teether} generates exploits for vulnerable contracts using symbolic execution with the Z3 constraint solver \cite{z3} to solve path constraints for the critical paths in the control flow graph. Contractfuzzer \cite{jiang2018contractfuzzer} uses the Application Binary Interface (ABI) specifications of vulnerable smart contracts to generate exploits (fuzzing inputs) for two vulnerabilities. SMARTSCOPY \cite{Feng2019smartscopy} also synthesizes adversarial contracts for exploiting vulnerabilities in contracts based on ABI specifications of the contracts covering larger set of vulnerabilities than Contractfuzzer. Echidna \cite{echidna}  has been proposed for fuzzing smart contracts. It supports grammar-based fuzzing to generate transactions to test smart contracts. The goal of these techniques is testing the smart contracts themselves for security vulnerabilities rather than testing bug-finding tools.
\pdfoutput=1

\section{Conclusion}
\label{sec:conclu}
This paper proposed \sysname, a technique for performing systematic evaluation of Ethereum smart contract's static analysis tools based on bug injection. \sysname analyzes the AST (Abstract Syntax Tree) of smart-contracts and injects pre-defined bug patterns at all possible locations in the AST.
\sysname was used to evaluate 6 smart contract static analysis tools, and the evaluation results show several cases of bugs that were not detected by the evaluated tools even though those undetected bugs are within the detection scope of tools. \sysname thus identifies important gaps in current static analysis tools for smart contracts, and provides a reproducible set of tests for developers of future static analysis tools. It also allows smart contract developers to understand the limitations of existing static analysis tools with respect to detecting security bugs. 

Future work will consist of expanding \sysname to other smart contract languages than Solidity, and automating the bug definition processes for injecting new bug types. 

\section{Acknowledgments}
	This work was partially supported by the Natural Sciences and Engineering Research Council of Canada (NSERC), and a research gift from Intel. We thank Julia Rubin, Sathish Gopalakrishnan, Konstantin Beznosov, and the anonymous reviewers of ISSTA'20 for their helpful comments about this work. 
%


\bibliographystyle{ACM-Reference-Format}
\bibliography{mainbib}


\end{document}